# TopoCompress: Topology Aware Token Compression Algorithm for Distributed Edge MoE Inference

Ning Li, Xinyu Wang, Xin Yuan, Wenchao Xu, Song Guo, *Fellow, IEEE*, Haijun Zhang, *Fellow, IEEE*

**Abstract—Mixture-of-experts (MoE) models increase capacity with moderate computational overhead by activating only a sparse subset of experts for each token. However, deploying MoE models on resource-constrained edge servers can cause substantial cross-server communication because activated experts are distributed across heterogeneous servers and network links. Existing expert placement and offloading methods mainly optimize deployment using raw token traffic, while conventional token compression focuses on semantic importance without considering topology-dependent routing costs. Thus, independently optimizing deployment and compression can result in inefficient communication and resource use. This paper proposes TopoCompress, a deployment- and topology-aware token compression framework for communication-efficient distributed edge MoE inference. TopoCompress jointly optimizes token compression, expert placement and replication, GPU-CPU residency, and collaborative token routing to balance cross-server traffic, inference quality, and deployment cost. To handle the coupling between token compression and deployment, it uses a two-timescale alternating optimization framework. The online fast loop compresses tokens with low semantic importance but high topology-induced routing costs and routes the remaining activations under the current deployment. The offline slow loop updates expert placement, replication, and GPU-CPU residency based on post-compression traffic. Simulations show that TopoCompress reduces cross-server traffic and resource consumption while maintaining controllable inference quality, enabling efficient distributed MoE inference on constrained edge infrastructures.**

## I. INTRODUCTION

Artificial intelligence (AI) has been widely used in intelligent assistants [1-2], autonomous driving [2-4], and content generation [5], significantly changing our lifestyle. However, high-performance AI models, especially large language models (LLMs), require substantial computing and memory resources. To improve model capacity without proportionally increasing computation, the Mixture-of-Experts (MoE) architecture has been widely adopted, where each token activates only a small subset of experts through Top-k gating [6-9]. Such sparse activation makes MoE models suitable for resource-limited edge environments, including mobile phones, vehicles, and unmanned aerial vehicles (UAVs).

One possible edge deployment solution is to distribute experts across edge servers and collaboratively execute the experts activated by each token, reducing the memory requirement of a single server and the inference latency [10-17]. Previous studies have investigated distributed deployment and collaborative inference among devices, edge servers, and clouds [18-25]. These studies optimize model placement, caching, or offloading to reduce inference latency and resource requirements while maintaining inference accuracy through activation-aware caching [18], expert placement optimization [19-21], and heuristic offloading [22-25].

However, these algorithms mainly optimize model deployment, caching, and offloading while overlooking token compression, another critical approach to reducing cross-server transmission. Although token compression has been extensively studied for LLM inference [26-33], existing methods cannot work effectively in distributed edge MoE scenarios because of the differences from single-machine inference [18-25], [34]. Methods such as FastGen, SnapKV, and Token Merging [26-30] prune or merge tokens based only on their model-level semantic importance and are unaware of how retained tokens affect cross-server routing frequency and communication costs. Moreover, the following issues in distributed edge MoE inference remain insufficiently addressed.

First, when experts are distributed and redundantly deployed across edge servers, the Top-k experts activated by a token may reside on different servers, requiring cross-server routing during layer-by-layer inference. Because tokens have heterogeneous gating distributions, the routing cost of each token depends jointly on its activation pattern and the current deployment topology. Thus, two tokens at the same layer and server may incur substantially different cross-server costs. Importantly, tokens that trigger expensive cross-server hops are not necessarily semantically unimportant. Therefore, compressing tokens that simultaneously have low semantic value and high routing cost can reduce the cross-server transmission that dominates inference latency in bandwidth-limited edge networks while preserving inference quality.

Second, existing works optimize expert deployment for raw token traffic [18-25] without considering how token compression reshapes the traffic actually transmitted. An optimal deployment for raw traffic is therefore not necessarily optimal for compressed traffic. For example, an expert replica may be deployed because a group of tokens frequently routes to it. If most of these tokens are

This paragraph of the first footnote will contain the date on which you submitted your paper for review, which is populated by IEEE. This work was supported in part by the grant from NSFC Grant no. 62571156, 62101159, NSF of Shandong Grant no. ZR2021MF055, the Research Grants Council of Hong Kong under the Areas of Excellence scheme grant AoE/E-601/22-R. *(Corresponding author: Xin Yuan).*

Ning Li and Haijun Zhang are with the School of Artificial Intelligence, University of Science and Technology Beijing, China (e-mail: ningli_polyuhk@outlook.com, zhanghaijun@ustb.edu.cn).

Xin Yuan is with the School of Ocean Engineering, Harbin Institute of Technology, Heilongjiang, China (e-mail: xin.yuan@hit.edu.cn).

Xinyu Wang is with the School of Computer Science and Technology, Harbin Institute of Technology, Heilongjiang, China (e-mail: wangxywy@gmail.com).

Wenchao Xu is with the Division of Integrative Systems and Design, Hong Kong University of Science and Technology, Hong Kong (e-mail: wenchaoxu@ust.hk)

Song Guo is with the Department of Computer Science and Engineering, Hong Kong University of Science and Technology, Hong Kong (e-mail: songguo@cse.ust.hk)

Mentions of supplemental materials and animal/human rights statements can be included here.

Color versions of one or more of the figures in this article are available online at http://ieeexplore.ieee.org

subsequently compressed because of their high routing costs and low importance, the replica continues occupying GPU memory without serving meaningful traffic. Conversely, replicas that continue serving frequent high-cost traffic after compression should be preserved. Hence, optimizing expert deployment only for raw traffic cannot guarantee resource efficiency because token compression and expert deployment are coupled.

These observations indicate that jointly optimizing token compression and expert deployment can reduce cross-server transmission and deployment resource consumption while maintaining high inference quality. Compressing high-cost, low-value tokens according to the deployment topology is particularly important when link bandwidth and GPU memory are limited. More importantly, token compression and expert deployment exhibit bidirectional coupling: the deployment topology determines which tokens should be compressed, while the compressed traffic determines how experts should be deployed and redundantly replicated.

Based on the above analysis, we propose TopoCompress, a Deployment- and Topology-Aware Token Compression algorithm for communication-efficient distributed edge MoE inference. TopoCompress jointly optimizes token compression, expert deployment, GPU-CPU residency, and token routing by considering cross-server transmission, inference quality, and deployment resource consumption. Since these objectives cannot be optimized simultaneously, we formulate a weight-based joint optimization problem to determine their optimal tradeoff. Furthermore, because token compression is performed per token while expert deployment is updated per epoch, we develop a two-timescale alternating optimization framework to address their coupling. The online fast loop performs topology-aware token compression and routing under the current deployment, while the offline slow loop re-optimizes expert deployment and redundant replication based on the accumulated post-compression traffic. We also investigate the feasibility, optimality, convergence, and computational complexity of the proposed algorithm.

The main contributions of this paper are summarized as follows.

- To the best of our knowledge, this is the first work to investigate deployment- and topology-aware token compression for distributed edge MoE inference. We formulate a joint optimization problem that balances cross-server transmission, inference quality, and deployment resource consumption by jointly determining token compression, expert deployment, GPU-CPU residency, and token routing strategies.
- Considering the bidirectional coupling between per-token compression and per-epoch deployment, we propose TopoCompress to improve joint optimization efficiency. Its online fast loop compresses high-cost, low-value tokens and collaboratively routes the surviving expert activations under the current deployment, while its offline slow loop re-optimizes expert deployment and redundant replication according to the accumulated post-compression traffic.
- We investigate the properties of TopoCompress. It always produces a quality-feasible solution under the per-token quality budget, and the number of participating servers per layer is bounded by the Top-k value. Moreover, its online routing overhead is independent of the number of edge servers, making token-level decisions practical for distributed edge MoE inference.

## II. Related Works

In this section, the works most related to TopoCompress are reviewed, and their limitations in distributed edge MoE inference are discussed.

### *A. Distributed and collaborative inference of edge LLMs*

Distributed and collaborative inference deploys large models across resource-constrained devices, edge servers, or cloud servers to reduce per-node resource burdens and inference latency [10–17]. Several surveys have summarized the convergence of edge computing and deep learning and the deployment of large models at the network edge [10], [11], [37], [42-43]. For dense neural networks, classical device-edge co-inference jointly optimizes model placement and online model splitting under fading wireless channels [35]. For LLMs, EdgeShard [1] collaboratively deploys model shards across heterogeneous devices by jointly optimizing device selection and model partitioning, while SplitLLM [36] places LLM sub-models across fog and edge-cloud nodes according to the model structure and input sequence length. Recent works, such as MDI-LLM [38], further enable distributed inference through device-to-device activation exchange. For MoE models, DanceMoE [19] accelerates inference through activation-aware, latency-optimized expert placement over heterogeneous GPU-equipped edge servers.

However, these works mainly target dense Transformers or block-wise partitioning with serial layer- or block-level inference paths. An MoE layer instead activates a sparse subset of experts, making the key decisions which tokens are worth the cross-server cost and how the retained tokens should be routed. Moreover, existing approaches transmit full token representations and cannot exploit the fact that much cross-server traffic is generated by tokens that are both semantically unimportant and costly under the current topology.

### *B. MoE serving, expert placement, and expert offloading*

MoE serving systems exploit expert sparsity and activation patterns to reduce memory and communication overhead through expert prediction, caching, prefetching, or offloading [18–21], [39–41]. MoE-Infinity [18] performs request-level, sparsity-aware expert caching and offloading between host and GPU memory, while SiDA-MoE [35] predicts expert activation to coordinate main and GPU memory. Pre-gated MoE [20] prefetches next-layer experts before gating to alleviate the CPU-offloading memory-performance tradeoff. Tutel [36], MegaBlocks [37], and Lina [38] improve MoE efficiency through adaptive expert parallelism, optimized all-to-all communication, or block-sparse computation. MoE-Prism [39] divides experts into fine-grained sub-experts to provide tunable throughput-latency tradeoffs, while DanceMoE [19] optimizes expert placement for edge environments.

However, these methods have several limitations in heterogeneous edge networks. First, many offloading systems focus on a single server or tightly coupled GPU-host memory hierarchies rather than heterogeneous Internet links. Second, expert placement is optimized for raw token traffic, whereas TopoCompress reshapes the cross-server traffic through token compression; consequently, a placement optimal for raw traffic may be inefficient for compressed traffic. Third, these methods treat deployment as the only optimization lever and do not compress the token stream itself. TopoCompress instead couples expert deployment and token compression bidirectionally: deployment determines which tokens are compressed, while the resulting traffic re-optimizes expert deployment and redundant replication.

*C. Token compression for efficient large model inference*

Token compression reduces the number or size of token representations involved in inference, thereby lowering computation and memory costs [26–30], [44–46]. One line of work prunes the KV cache according to token importance. H2O [44] retains heavy-hitter tokens identified by accumulated attention scores; StreamingLLM [45] preserves attention sinks and a sliding window; FastGen [26] selects per-head compression policies through offline profiling; SnapKV [27] identifies salient KV positions from a prompt-end observation window; and PyramidKV [46] allocates layer-wise KV budgets according to the pyramidal distribution of attention information. Another line merges similar tokens to shorten sequences, as exemplified by ToMe [28], which performs bipartite soft matching without retraining. Surveys have also summarized efficient LLM inference and compression techniques, including quantization, pruning, distillation, and KV-cache optimization [20], [47], [48].

However, these methods target single-machine or model-level inference and determine compression priorities solely from semantic importance, attention magnitude, or token similarity. They are unaware of how retained tokens affect cross-server routing and communication costs in distributed edge MoE systems. Consequently, they may retain expensive remotely routed tokens or compress local tokens that offer little communication saving. Moreover, compressed traffic is not fed back to expert deployment, leaving compression and deployment disconnected. TopoCompress addresses this gap through a deployment-topology-aware, dual-dimensional compression score that jointly considers semantic importance and topology-induced routing cost, and closes the loop between token compression and expert deployment through two-timescale alternating optimization.

## III. Network Model and Problem Statement

In this section, the token semantic importance model, the topology-aware routing cost model, the token compression and quality degradation model, and the per-task quality budget, which are used in this paper are introduced in detail.

*A. Network Model and Distributed MoE Deployment*

This paper introduces a distributed edge intelligence system comprising a set of geographically distributed edge servers $\mathcal{N} = \{1, 2, \dots, N\}$ and a set of mobile users $\mathcal{U} = \{1, 2, \dots, U\}$. The edge servers are interconnected through ordinary Internet links with heterogeneous and time-varying characteristics. Let $B_{m,n}$ and $d_{m,n}$ denote the available bandwidth and the propagation-plus-queuing delay between server $s_m$ and $s_n$, respectively, both of which differ across server pairs. Each server $s_n \in \mathcal{N}$ is characterized by its GPU memory capacity $G_n^M$, CPU memory capacity $G_n^C$, GPU computation capability $F_n$, and GPU-CPU transfer bandwidth $\beta_n$. For each user $i \in \mathcal{U}$, its associated entry server is denoted as $a_i \in \mathcal{N}$, which receives the inference request and serves as the initial token execution server. Unlike conventional designs in which a fixed server aggregates every layer, in this paper the residing server of a token evolves dynamically along its layer-by-layer trajectory.

The MoE model consists of $L$ layers. Each layer $l \in \{1,2,\dots,L\}$ contains a set of experts $\mathcal{E}_l$, among which the Top-$k$ gating mechanism activates a subset $\mathcal{K}_{i,j,l} \subseteq \mathcal{E}_l$ with $|\mathcal{K}_{i,j,l}| = k$ for each input token $\tau_{i,j}$. Since for the distributed deployment of edge LLMs, such as [23,40], the activated experts may reside on different edge servers, each token may require cross-server transmission during its layer-by-layer inference process, leading to significant communication overhead under limited and heterogeneous edge link bandwidth.

The expert deployment strategy is characterized by the binary variable $x_{E,n} \in \{0,1\}$, where $x_{E,n} = 1$ indicates that expert $E$ is deployed on server $s_n$. Let $\boldsymbol{X} = \{x_{E,n}\}$ denote the full deployment matrix. The GPU residency variable $y_{E,n} \in \{0,1\}$ indicates whether the deployed replica of $E$ on $s_n$ is loaded into GPU memory ($y_{E,n} = 1$) or kept in CPU memory ($y_{E,n} = 0$ with $x_{E,n} = 1$); $\boldsymbol{Y} = \{y_{E,n}\}$ denotes the residency matrix. A CPU-resident replica must be loaded into GPU memory before execution, which incurs an offload delay. This is the mechanism through which $\boldsymbol{Y}$ affects the realized execution cost, made precise in (2a) below. Each replica carries a quantization level $c_{E,n}$, with $c_{max}$ denoting full precision. To guarantee inference correctness, we adopt the full-precision invariant:

$$\exists n \in \mathcal{N}: x_{E,n} = 1 \wedge c_{E,n} = c_{max}, \forall E \in \mathcal{E}_l, \forall l \tag{1}$$

For user $i$, its token sequence is denoted as $\mathcal{T}_i = \{\tau_{i,1}, \tau_{i,2}, \dots\}$. The token $\tau_{i,j}$ at layer $l$ resides on server $\theta_{i,l} \in \mathcal{N}$, which evolves dynamically according to the routing decisions made at each layer, with $\theta_{i,1} = a_i$ as the sole fixed point. The cross-server communication cost incurred by routing token $\tau_{i,j}$ from $\theta_{i,l}$ to a remote server $s_n$ hosting an activated expert is expressed as:

$$T_{\theta_{i,l},n}^{\text{comm}} = \frac{v_{i,j,l}}{B_{\theta_{i,l},n}} + d_{\theta_{i,l},n} \tag{2}$$

where $v_{i,j,l}$ denotes the volume of the token representation transmitted at layer $l$. When the selected replica of $E$ on $s_n$ is CPU-resident, an additional GPU-CPU offload delay is incurred before execution:

$$T_{E,n}^{\text{off}} = (1 - y_{E,n}) \frac{m_E}{\beta_n} \tag{2a}$$

where $m_E$ is the memory footprint of expert $E$; so that a locally available but CPU-resident replica may be slower than a remote but GPU-resident one. It is worth noting that $v_{i,j,l}$ in (2) is directly determined by the token compression decision applied at or before layer $l$, which constitutes the central coupling between the token compression model and the routing cost model. The transmission term in (2) vanishes when the activated expert is co-located with the residing server $\theta_{i,l}$. Consequently, executing a token locally is a zero transmission candidate rather than a default optimum, and the routing choice depends jointly on the deployment topology, the GPU residency, and the compression decision.

**Scope of the cost model.** This paper deliberately targets the communication bottleneck of distributed edge MoE inference, so the objective that will be presented in Section III.F is formulated over cross-server transmission, quality degradation, and deployment memory. The GPU computation delay, which scales with the expert load and the server capability $F_n$, is not placed in the objective for two reasons: first, it is largely orthogonal to the token compression decision that this paper optimizes, since compressing a token reduces transmitted volume rather than per-expert FLOPs; second, the placement-side computation-latency trade-off is studied in our companion distributed-routing work [41]. The capability $F_n$ is therefore retained and used only in the online per-server stability guard introduced in Section III.F, which prevents a server from being overloaded, while the optimization focus remains communication.

*B. Token Semantic Importance Model*

Different tokens contribute unequally to the final inference output. Tokens that carry rich semantic content or are critical to downstream expert activations should be preserved with high fidelity, while tokens with redundant representations or low-discriminative representations can tolerate aggressive compression. Consistent with established token pruning literature [Y], we characterize the semantic importance of token $\tau_{i,j}$ at layer $l$ by its normalized attention accumulation score:

$$\mathrm{Imp}(\tau_{i,j},l)=\frac{\sum_{h=1}^{H}\sum_{t=1}^{T}A_{t,j,l}^{(h)}}{\sum_{j'=1}^{T}\sum_{h=1}^{H}\sum_{t=1}^{T}A_{t,j',l}^{(h)}} \quad (3)$$

where $A_{t,j,l}^{(h)}$ denotes the attention weight that query position $t$ assigns to token $\tau_{i,j}$ (at key position $j$) in head $h$ at layer $l$, $H$ is the number of heads, and $T$ is the current sequence length. Here $t$ indexes the query position and $j$ indexes the key position, so the inner sum over $t$ accumulates how strongly token $j$ is attended to by all queries. This column-sum convention is the standard attention-based importance measure. By construction, $\mathrm{Imp}(\tau_{i,j},l)\in[0,1]$ and $\sum_j \mathrm{Imp}(\tau_{i,j},l)=1$.

A token with a low importance score makes a disproportionately small contribution to the output representation and is therefore a candidate for compression. However, semantic importance alone is insufficient to determine compression priority in the distributed edge MoE setting, because it does not capture the communication consequences of retaining or removing a token. This observation motivates the topology aware routing cost model introduced in the following subsection.

*C. Topology Aware Routing Cost Model*

Given the deployment matrix $\boldsymbol{X}$ and the token's current residing server $\theta_{i,l}$, we measure the per-hop communication overhead by a unit-volume link cost that depends only on the link characteristics and is independent of the token volume. Concretely, for each activated expert $E\in\mathcal{K}_{i,j,l}$, let $\mathcal{C}_E^{ex}=\{(E,n)|x_{E,n}=1\}$ denote the set of servers hosting a replica of expert $E$, and define the minimum unit-volume routing cost from the current server $\theta_{i,l}$ is:

$$\Gamma(E,\theta_{i,l},\boldsymbol{X})=\min_{n:x_{E,n}=1}\left(\frac{1}{B_{\theta_{i,l}n}}+d_{\theta_{i,l}n}\right) \quad (4)$$

By using the unit-volume link cost rather than $T^{\mathrm{comm}}$ of (2), $\Gamma$ reflects only the link characteristics induced by the deployment topology and does not depend on the token volume $v_{i,j,l}$ or on the compression ratio $\rho_{i,j,l}$. The effect of token volume and compression is deferred to the post-compression transmission delay in (8). This removes the circular dependence that would otherwise arise if the routing-cost graph were built from a volume dependent cost while being used to decide the very compression that changes that volume. The topology aware routing cost of token $\tau_{i,j}$ aggregated over its activated experts at layer $l$ is then:

$$\mathcal{C}(\tau_{i,j},l,\boldsymbol{X})=\frac{1}{k}\sum_{E\in\mathcal{K}_{i,j,l}}\Gamma(E,\theta_{i,l},\boldsymbol{X}) \quad (5)$$

It is noteworthy that $\mathcal{C}(\tau_{i,j},l,\boldsymbol{X})$ depends on both the deployment topology $\boldsymbol{X}$ and the specific expert activation pattern of the token. Consequently, two tokens at the same layer and on the same server may have substantially different routing costs if their gating distributions differ, which is precisely the heterogeneity that existing topology independent compression methods fail to exploit.

The routing cost in (5) is evaluated with the gating distribution observed before compression is applied at layer $l$. Because compression perturbs the hidden state of a token, its Top-$k$ activation set $\mathcal{K}_{i,j,l}$ may change after compression, so the pre-compression routing cost is in principle an estimate of the post-compression one. In this paper, we adopt the pre-compression gating distribution as a first-order estimator or the routing cost, and we empirically verify in Section VI that the effect of compression on the shallow layer gating decision is small, so that this estimator is accurate enough to drive the compression priority. This assumption is stated explicitly here rather than left implicit, since it directly affects the fidelity of (5).

To enable efficient online computation, $\mathcal{C}(\tau_{i,j},l,\boldsymbol{X})$ is approximated via a precomputed routing cost graph $\mathcal{G}=\{\Gamma(E,n,\boldsymbol{X})\}_{E,n}$ during the offline phase. For each expert $E$ and each server $s_n$, the pairwise minimum unit-volume routing cost $\Gamma(E,n,\boldsymbol{X})$ is precomputed from the current deployment $\boldsymbol{X}$ and the inter-server link parameters $\{B_{m,n},d_{m,n}\}$. During online inference, the per-token routing cost is obtained by a lightweight lookup into $\mathcal{G}$ indexed by the token's current server $\theta_{i,l}$ and its Top-$k$ activated expert set, incurring negligible scheduling overhead. Because $\boldsymbol{X}$ is itself refreshed by the offline stage according to the post-compression traffic, the cost graph $\mathcal{G}$ is recomputed once per deployment epoch rather than per token.

Building upon (3) and (5), the dual-dimensional compression score of token $\tau_{i,j}$ at layer $l$ is:

$$\mathcal{S}(\tau_{i,j},l)=\mathrm{Imp}(\tau_{i,j},l)\times\mathcal{C}(\tau_{i,j},l,X) \quad (6)$$

The rationale underlying (6) is as follows. A token with a low importance score contributes little to quality, while a token with a high routing cost triggers expensive cross-server transmission. Consequently, a token yielding a low dual-dimensional score $\mathcal{S}(\tau_{i,j},l)$ is simultaneously low in semantic value and high in communication overhead, making it the most cost-effective target for aggressive compression. Conversely, tokens with high scores are preserved with full precision to protect both inference quality and routing efficiency. We emphasize that (6) collapses to a purely semantic criterion when the routing cost is identical across tokens, i.e., when the deployment topology induces no heterogeneity; in that degenerate case TopoCompress reduces to a conventional topology independent compression scheme, which it therefore strictly generalizes.

*D. Token Compression Model and Quality Degradation*

Token compression is realized by two complementary discrete operations: 1) importance-based pruning, which removes a token whose dual-dimensional score falls below a threshold, and 2) similarity-based merging, which fuses a group of similar tokens into a single representative token. Both operations reduce the number of token representations transmitted across servers. To express these discrete operations within a single tractable variable, we introduce the per-token effective compression ratio $\rho_{i,j,l}\in[0,1]$ at layer $l$, where $\rho_{i,j,l}=1$ indicates no compression and $\rho_{i,j,l}=0$ indicates full removal. Pruning corresponds to $\rho_{i,j,l}\in\{0,1\}$, whereas merging corresponds to an intermediate value equal to the fraction of the original volume retained by the merged representative. Thus, $\rho_{i,j,l}$ is the continuous relaxation of the underlying discrete compression operations, and the online algorithm in Section V will round it back to a feasible prune-or-merge action. The effective transmission volume of token $\tau_{i,j}$ at layer $l$ is then:

$$v_{i,j,l}^{\mathrm{comp}}=\rho_{i,j,l}\cdot v_{i,j,l} \quad (7)$$

where $v_{i,j,l}$ denotes the uncompressed token representation size in bytes. It follows directly from (7) and (2) that the cross-server communication delay after compression is:

$$T^{\text{comm,comp}}_{\theta_{i,l},n} = \frac{\rho_{i,j,l} v_{i,j,l}}{B_{\theta_{i,l},n}} + d_{\theta_{i,l},n} \quad (8)$$

Hence, reducing $\rho_{i,j,l}$ directly reduces the per-hop transmission time, with the savings proportional to the uncompressed volume $v_{i,j,l}$ and concentrated on high-cost remote hops by the score in (6).

Compression inevitably introduces quality degradation into the inference output. Let $\delta_l(\rho_{i,j,l})$ denote the per-token quality degradation incurred by compressing token $\tau_{i,j}$ to ratio $\rho_{i,j,l}$ at layer $l$. We model $\delta_l(\cdot)$ as a monotonically decreasing and convex function of $\rho_{i,j,l}$, calibrated offline via empirical profiling on a representative calibration dataset, with $\delta_l(1) = 0$ and $\delta_l(0) = \delta_{max,l}$. A common parametric form adopted in this paper is:

$$\delta_l(\rho_{i,j,l}) = \delta_{max,l}(1 - \rho_{i,j,l})^{\gamma_l} \quad (9)$$

where the shape parameter $\gamma_l > 0$ and the maximum loss $\delta_{max,l}$ are calibrated per layer rather than assumed uniform, since the sensitivity of compression varies across depth: shallow layers, whose representations are less stabilized, are typically more sensitive than deep layers. The accumulated quality degradation for user $i$ across all tokens and all layers is:

$$\Lambda^{\text{acc}}_i = \sum_{l=1}^{L} \sum_{j=1}^{|\mathcal{T}_i|} \delta_l(\rho_{i,j,l}) \quad (10)$$

The (10) is stated as a first-order additive surrogate of the end-to-end quality loss rather than an exact decomposition. Token compression at one layer affects the attention and gating of subsequent layers, so the per-layer degradations are not strictly independent. The (10) is adopted because it is additive and tractable for the budget constraint, and its correlation with the measured end-to-end task metric is validated empirically in Section VI.

*E. Task Quality Budget*

Different users has different requirements on inference quality. Even for the same user, the tolerable quality degradation varies across tasks and input contexts. In this paper, we adopt the accumulated quality degradation $\Lambda^{\text{acc}}_i$ defined in (10) as the task-level quality metric. Each user $i$ is associated with a per-task quality budget $\Lambda^{max}_i$, the maximum tolerable cumulative degradation.

**Definition 1 (Task Quality Budget).** The task quality budget $\Lambda^{max}_i$ is the maximum accumulated quality degradation that user $i$ can tolerate while maintaining the task-level inference accuracy above a prescribed threshold $\epsilon_i$. The compression strategy is quality-feasible for user $i$ if and only if $\Lambda^{\text{acc}}_i \le \Lambda^{max}_i$.

The parameter $\Lambda^{max}_i$ is calibrated offline by profiling the target model on task-specific evaluation datasets. For a given application (e.g., question answering, code generation, or summarization), it is determined as the largest cumulative degradation level at which task accuracy remains within $\epsilon_i$ of the uncompressed baseline. The mapping from $\epsilon_i$ to $\Lambda^{max}_i$ is obtained from a profiled degradation-accuracy curve, whose reliability will be reported in Section VI.

It is worth emphasizing that, unlike latency based QoE models that treat all users uniformly, the quality budget framework is inherently heterogeneous: users running latency insensitive, quality critical tasks (e.g., medical report generation) are assigned tight budgets $\Lambda^{max}_i \approx 0$, or even $\Lambda^{max}_i = 0$ to forbid any compression, while users on latency sensitive, quality tolerant tasks (e.g., real time dialogue) are assigned looser budgets. This heterogeneity enables TopoCompress to compress only those tokens for which quality slack exists.

*F. Problem Statement*

As established in Sections III.A through III.E, the cross-server communication incurred during the inference of user $i$'s token sequence is jointly determined by four coupled factors: the deployment matrix $\boldsymbol{X}$, the residency matrix $\boldsymbol{Y}$, the compression matrix $\boldsymbol{\rho}$, and the routing strategy $\boldsymbol{Z}$. The total post-compression cross-server transmission volume for user $i$ is:

$$V_i(\boldsymbol{X}, \boldsymbol{\rho}, \boldsymbol{Z}) = \sum_{l=1}^{L} \sum_{j=1}^{|\mathcal{T}_i|} \sum_{E \in \mathcal{K}_{i,j,l}} \sum_n z_{i,j,l,E,n} v^{\text{comp}}_{i,j,l} \quad (11)$$

where $z_{i,j,l,E,n} \in \{0,1\}$ is the routing indicator with $z_{i,j,l,E,n} = 1$ indicating that expert $E$ for token $\tau_{i,j}$ at layer $l$ is executed on server $s_n$, and $\sum_n z_{i,j,l,E,n} = 1$ for each $(i, j, l, E)$. The corresponding total cross-server communication delay for user $i$ at layer $l$ is:

$$T^{\text{comm}}_{i,l}(\boldsymbol{X}, \boldsymbol{Y}, \boldsymbol{\rho}, \boldsymbol{Z}) = \sum_{j=1}^{|\mathcal{T}_i|} \sum_{E \in \mathcal{K}_{i,j,l}} \sum_n z_{i,j,l,E,n} \left( T^{\text{comm,comp}}_{\theta_{i,l},n} + T^{\text{off}}_{E,n} \right) \quad (12)$$

which includes the GPU-CPU offload delay $T^{\text{off}}_{E,n}$ of (2a), so that the residency matrix $\boldsymbol{Y}$ explicitly influences the realized delay. A key point that distinguishes this formulation from a one-directional design is that the deployment is evaluated against the post-compression traffic rather than the raw traffic. To make this reverse coupling explicit, we define the traffic aware deployment cost on server $s_n$ as the memory footprint of its replicas minus the deployment value they provide under the compressed traffic:

$$M_n(X, \boldsymbol{\rho}, Z) = \sum_E x_{E,n} m_E - \eta \sum_E x_{E,n} \Phi_{E,n}(\boldsymbol{\rho}, \boldsymbol{Z}) \quad (13)$$

where $m_E$ is the memory footprint of expert $E$, $\Phi_{E,n}(\boldsymbol{\rho}, \boldsymbol{Z})$ is the post-compression cross-server traffic that the replica of expert $E$ on server $s_n$ actually serves, and $\eta \ge 0$ weights the traffic-relief value against memory price. The second term is the mechanism through which compression feeds back into deployment, i.e., after compression, a replica that no longer serves meaningful cross-server traffic contributes little value and is therefore not worth its memory, whereas a replica that still serves frequent high-cost traffic is rewarded. Thus $\boldsymbol{\rho}$ enters the deployment objective directly, closing the compression-to-deployment loop.

To combine the three objectives, which are of different natures and units, we normalize each by its reference value under the uncompressed full-precision strategy on the same deployment, so that the objective is dimensionless and additive. Let $V^{ref}$, $\Lambda^{max}$, and $M^{ref}$ denote the reference transmission volume, the per-user maximum tolerable degradation, and the reference deployment memory, respectively. The optimization problem P0 is then:

$$\min_{\boldsymbol{X},\boldsymbol{Y},\boldsymbol{\rho},\boldsymbol{Z}} \sum_{i \in \mathcal{U}} \frac{V_i(\boldsymbol{X},\boldsymbol{\rho},\boldsymbol{Z})}{V^{\text{ref}}_i} + \mu \sum_{i \in \mathcal{U}} \frac{\Lambda^{\text{acc}}_i(\boldsymbol{\rho})}{\Lambda^{max}_i} + \nu \sum_{i \in \mathcal{U}} \frac{M_n(\boldsymbol{X},\boldsymbol{\rho},\boldsymbol{Z})}{M^{\text{ref}}_n} \quad (14)$$

subject to:

$$\Lambda^{\text{acc}}_i(\boldsymbol{\rho}) \le \Lambda^{max}_i, \forall i \in \mathcal{U} \quad \text{(c.1)}$$
$$\sum_E x_{E,n} y_{E,n} m_E \le G^M_n, \forall n \in \mathcal{N} \quad \text{(c.2a)}$$
$$\sum_E x_{E,n}(1 - y_{E,n}) m_E \le G^C_n, \forall n \in \mathcal{N} \quad \text{(c.2b)}$$
$$y_{E,n} \le x_{E,n}, \forall E, \forall n \quad \text{(c.2c)}$$
$$\sum_{n \in \mathcal{N}} x_{E,n} \ge 1, \forall E \in \mathcal{E}_l, \forall l \quad \text{(c.3)}$$
$$\exists n: x_{E,n} = 1 \ \wedge \ c_{E,n} = c_{max}, \forall E \in \mathcal{E}_l, \forall l \quad \text{(c.4)}$$
$$\sum_n z_{i,j,l,E,n} = 1, \forall i, j, l, E \in \mathcal{K}_{i,j,l} \quad \text{(c.5)}$$
$$z_{i,j,l,E,n} \le x_{E,n}, \forall i, j, l, E, n \quad \text{(c.6)}$$
$$0 \le \rho_{i,j,l} \le 1, \forall i, j, l \quad \text{(c.7)}$$
$$x_{E,n}, y_{E,n}, z_{i,j,l,E,n} \in \{0,1\}, \forall E, n, i, j, l \quad \text{(c.8)}$$

In Problem P0, the objective is dimensionless: each term is divided by its reference value, so the weights $\mu$ and $\nu$ are unitless preference weights that price quality degradation and traffic-aware deployment cost against the normalized transmission volume; their sensitivity is reported in Section VI. Constraint (c.1) enforces the per-user task quality budget as a hard QoS bound. Constraints (c.2a) and (c.2b) split the per-server memory budget into a GPU-resident part and a CPU-resident part, so that $\boldsymbol{Y}$ consumes a concrete resource. Constraint (c.2c) ensures a replica can reside in GPU memory only if it is deployed. Constraint (c.3) guarantees each expert has at least one deployed replica, and (c.4) enforces the full-precision invariant. Constraints (c.5) and (c.6) govern routing feasibility, and (c.7) to (c.8) specify the variable domains. The online stage additionally enforces a per-server stability guard using $F_n$, which is the only place $F_n$ is used.

Addressing Problem P0 is challenging because it couples three decision layers, which are: token compression (continuous), expert deployment (binary), and online routing (binary), across two-time scales. Because the routing cost depends on both $F_n$ and the online activation pattern, and the deployment value $\Phi_{E,n}(\boldsymbol{\rho}, \boldsymbol{Z})$ depends on compression, so the two directions of the coupling are mutually dependent. Additionally, since $\delta_l(\cdot)$ is nonlinear, making (c.1) non-convex. P0 therefore serves as a non-causal ideal benchmark and is decomposed into the offline deployment stage (Section IV) and the online compression-and-routing stage (Section V), which alternate on two-time scales.

## IV. Traffic-Aware Offline Deployment

In this section, we present the offline deployment scheme that determines the expert placement strategy $\boldsymbol{X}$, the GPU-CPU residency strategy $\boldsymbol{Y}$, and the routing-cost graph $\mathcal{G}$ consumed by the online stage.

The offline deployment process runs once during each deployment period, using a separate calibration dataset while keeping the main system unchanged. The aim is not just to find any workable placement, but to carefully organize the deployment so that, when the online stage creates compressed traffic, costly cross-server transfers are avoided in the most important situations. To make the coupling explicit, every offline benefit term is defined as an expectation of the same normalized objective (14), evaluated over the calibration traffic under the online compression-and-routing policy that presented in Algorithm 2 (Section V, Table II).

Before the placement decision, we profile three quantities on the calibration token set $\mathcal{T}_{\text{cal}}$ under the current deployment and the online policy of Algorithm 2.

First, the post-compression activation traffic of an (expert, server) pair is:

$$\Phi_{E,n} = \frac{1}{|\mathcal{T}_{\text{cal}}|} \sum_{\tau \in \mathcal{T}_{\text{cal}}} \sum_l z_{\tau,l,E,n} v_{\tau,l}^{\text{comp}} \tag{15}$$

which measures how much cross-server traffic the replica of $E$ on $s_n$ actually serves after the online compression has removed the low value and high cost tokens. Unlike a raw activation frequency, $\Phi_{E,n}$ is computed on the compressed token stream, so it already reflects the reverse coupling of (13).

Second, for each candidate (expert, server) pair we define the GPU-residency benefit, weighted by how often the replica is actually used after compression:

$$g_{E,n} = \pi_{E,n}^{\text{sel}} \left(\frac{m_E}{\beta_n}\right) \tag{16}$$

where $\pi_{E,n}^{\text{sel}}$ is the empirical probability that, given $E$ is activated, the online router selects the replica on $s_n$, obtained by simulating Algorithm 2 on $\mathcal{T}_{\text{cal}}$. The factor $m_E / \beta_n$ is exactly the offload term $T_{E,n}^{\text{off}}$ of (2a) eliminated once the replica becomes GPU-resident.

Third, for the redundant-replication stage we define the redundancy benefit of adding a replica of $E$ on a new server $s_n$ is:

$$b_{E,n} = \mathbb{E}_{\tau \in \mathcal{T}_{\text{cal}}} \left[ \frac{(\hat{C}_E - \Gamma(E,n,X))^+ \overline{v}_{\text{comp}}}{V^{\text{ref}}} \right] + \eta \frac{\widehat{\Phi}_{E,n}}{M^{\text{ref}}} \tag{17}$$

where $\hat{C}_E$ is the current minimum unit-volume routing cost of $E$ over its existing candidate servers, $\Gamma(E, n, X)$ is the unit-volume cost of the new path in (4), $\overline{v}_{\text{comp}}$ is the average post-compression token volume, $(\cdot)^+ = \max\{\cdot, 0\}$, and $\widehat{\Phi}_{E,n}$ is the post-compression traffic the new replica is expected to relieve. The first term is the transmission saving of the new path normalized by $V^{\text{ref}}$; the second is its traffic-relief value normalized by $M^{\text{ref}}$, mirroring the value term of (13). A larger $b_{E,n}$ therefore means replicating $E$ on $s_n$ creates a cheaper path precisely for traffic that survives compression.

The details of Algorithm 1 are presented in Table I and explained as follows.

Table I

**Algorithm 1: Traffic-aware offline deployment**

**Input:** edge server set $\mathcal{N}$; calibration set $\mathcal{T}_{\text{cal}}$; baseline deployment; GPU/CPU memory $G_n^M$, $G_n^C$; link parameters $\{B_{m,n}, d_{m,n}\}$; GPU-CPU transfer bandwidth $\beta_n$.

**Output:** Deployment **X**, residency **Y**, routing-cost graph $\mathcal{G}$.

1. Initialize **X** from baseline satisfying the full-precision invariant;
2. set $y_{E,n} \leftarrow 0$
3. Simulate Algorithm 2 on $\mathcal{T}_{\text{cal}}$ under $(\boldsymbol{X}, \boldsymbol{Y})$ to obtain $\pi_{E,n}^{sel}$ and post-compression traffic $\Phi$ by (15)
4. for each server $s_n \in \mathcal{N}$ do
5. Build $\mathcal{A}_n = \{E | x_{E,n} = 1\}$;
6. compute $g_{E,n}$ by (16) for all $E$ in $\mathcal{A}_n$
7. Sort $\mathcal{A}_n$ descending by $g_{E,n}$;
8. set residual GPU memory $\tilde{G}_n \leq G_n^M$
9. for each E in $\mathcal{A}_n$ in sorted order do
10. if $m_E \leq \widehat{\mathrm{G}}_n$ then $y_{E,n} \leftarrow 1$; $\tilde{G}_n \leftarrow \tilde{G}_n - m_E$
11. end for
12. end for
13. while some server has free memory and a feasible new replica exists do
14. Compute $b_{E,n}$ by (17) for all feasible $(E, n)$ with $x_{E,n} = 0$, subject to (c.1) to (c.4)
15. Select $(E^*, n^*) = \arg\max b_{E,n}$
16. if $m_{E^*} \leq \widehat{\mathrm{G}}_{n^*}$ then
17. $x_{E^*,n^*} \leftarrow 1$; $y_{E^*,n^*} \leftarrow 1$; $\tilde{G}_{n^*} \leftarrow \tilde{G}_{n^*} - m_{E^*}$
18. else $x_{E^*,n^*} \leftarrow 1$; $y_{E^*,n^*} \leftarrow 0$; $\tilde{G}_{n^*}^C \leftarrow \tilde{G}_{n^*}^C - m_{E^*}$
19. end if
20. Update $\hat{C}_{E^*}$ and P $\Phi$ for the affected expert
21. end while
22. Re-run Stage 1 globally over the enlarged replica set with a diminishing-returns guard
23. Build routing-cost graph $\mathcal{G} = \{\Gamma(E, n, \boldsymbol{X})\}$ by (4) using final **X** and $(B_{m,n}, d_{m,n})$
24. return **X**, **Y**, $\mathcal{G}$

Profiling (lines 1-2): The algorithm starts from a baseline deployment that already satisfies the full-precision exact-replica invariant (c.4), which guarantees that, regardless of all later decisions, every expert remains reachable at full precision and the online fallback can never become quality infeasible. All replicas are initialized as CPU-resident, i.e. $y_{E,n} \leftarrow 0$. Algorithm 2 is then simulated on the calibration traffic under this initial $(\boldsymbol{X}, \boldsymbol{Y})$ to obtain two statistics that drive every subsequent decision: the empirical selection probability $\pi_{E,n}^{sel}$, i.e. how often the online router actually chooses the replica of $E$ on $s_n$ once low-value high-cost tokens have been compressed away, and the post-compression activation traffic $\Phi_{E,n}$ of (15). Profiling under the compressed stream is the key difference from a frequency-based scheme, i.e., a replica that looks hot on the raw token stream, but whose traffic is mostly removed by compression will receive a small $\pi_{E,n}^{sel}$ and a small $\Phi_{E,n}$, and will therefore not be over-provisioned.

Stage 1: routing-aware GPU-residency assignment (lines 3-10): This stage decides which replicas are promoted from CPU into the scarce GPU memory. For each server $s_n$, the algorithm forms the set $\mathcal{A}_n = \{E | x_{E,n} = 1\}$ of its deployed replicas and scores each by the GPU-residency benefit $g_{E,n} = \pi^{sel}_{E,n}(m_E/\beta_n)$ of (16). The factor $m_E/\beta_n$ is exactly the GPU-CPU offload delay, $T^{\text{off}}_{E,n}$ that is eliminated once the replica becomes GPU-resident, and the weight $\pi^{sel}_{E,n}$ discounts it by how often that saving is actually realized online. The replicas of $s_n$ are sorted in descending order of $g_{E,n}$ and greedily promoted into GPU memory until the GPU budget $G^M_n$ of constraint (c.2a) is exhausted. As a result, GPU memory is spent first on the replicas that are both expensive to offload and frequently selected, which directly removes the offload term that the online stage would otherwise pay in (18). A replica that is rarely selected, or that sits behind a slow link so the router seldom routes to it, obtains a small $g_{E,n}$ and is left in CPU memory.

Stage 2: traffic-aware redundant replication (lines 11-19): While free memory remains and a feasible new replica exists, this stage increases the candidate collaboration domain by inserting redundant replicas where they relieve the most surviving cross-server cost. In each iteration it evaluates the redundancy benefit $b_{E,n}$ of (17) for every feasible pair $(E, n)$ with $x_{E,n} = 0$ those respects (c.1) to (c.4) and inserts the single most beneficial pair $(E^*, n^*) = \arg\max b_{E,n}$ . Because $b_{E,n}$ combines the transmission saving of the new path, $\left(\hat{C}_E - \Gamma(E, n, \boldsymbol{X})\right)^+$, with the post-compression traffic $\widehat{\Phi}_{E,n}$ that the new replica is expected to relieve, replication is steered toward exactly those experts and servers where the cost that survives compression is highest, for example a hot expert whose only current replica sits behind a congested link. The new replica is made GPU-resident when GPU memory allows, since a GPU-resident alternative is what creates a genuinely low-cost online path. Otherwise, it falls back to a CPU-resident replica that still adds a candidate path while respecting the CPU budget (c.2b). After each insertion the minimum routing cost $\hat{C}_{E^*}$ and the traffic estimate $\Phi$ of the affected expert are updated, so the next iteration sees the diminished marginal benefit of further replicating the same expert and naturally spreads replicas across experts and servers.

Stage 3: residency re-optimization (lines 20-24): Inserting new replicas changes the memory landscape, so freezing the Stage 1 residency would leave **X** and **Y** mutually inconsistent. Stage 3 therefore re-runs the residency assignment globally over the enlarged replica set, under a diminishing-returns guard: a second GPU-resident copy of the same expert on a different server is admitted only if its marginal $g$ still exceeds the marginal of the best not-yet-resident distinct expert. This prevents the system from wasting GPU memory on duplicate GPU copies of one popular expert while a distinct, frequently selected, CPU-only expert starves. Finally, the routing-cost graph $\mathcal{G} = \{\Gamma(E, n, \boldsymbol{X})\}$ is built from the final deployment using (4), so that the online stage can read any per-hop unit-volume cost in $O(1)$. Because every benefit term in this algorithm is an expectation of the same normalized online objective (14) evaluated on the post-compression traffic $\Phi$ , the returned triple $(\boldsymbol{X}, \boldsymbol{Y}, \mathcal{G})$ is the deployment that minimizes the expected online routing cost under the current compression policy, which is precisely the reverse direction of the two-timescale loop: compression shapes the traffic, and the traffic shapes the deployment.

**Offline complexity.** Stage 1 costs $O(\sum_n |A_n| \log|A_n|$ to score and sort the residency candidates; Stage 2 costs $O(R|\mathcal{E}|N)$ over the calibration traffic, where $R$ is the replication cap; Stage 3 repeats the Stage 1 scan. Since this procedure runs once per deployment epoch rather than per token, its cost is fully amortized and adds nothing to the per-token online latency.

## V. Online Topology-Aware Compression and Collaborative Routing

The online stage processes tokens layer by layer over the deployment produced by Algorithm 1. For each token-layer, it first computes the dual-dimensional compression score (6), applies a compression action under the per-token quality budget, and then routes the surviving activated experts as a set by minimizing the post-compression bottleneck cost. The cost graph $\mathcal{G}$ makes the routing-cost lookup negligible, so the per-token overhead is dominated by the small set-level routing search.

### *A. Online Decision State and Per-Step Cost*

For token $\tau_{i,j}$, let $\theta_{i,l}$ denote the residing server before layer $l$, initialized to the home server $a_i$. After a layer is computed on the participating set $\mathcal{P}_{i,j,l}$, the partial outputs are gathered to an aggregation server that becomes the residing server of the next layer, so consecutive layers on the same server incur no transmission. At layer $l$, the gating network produces the Top-$k$ set $\mathcal{K}_{i,j,l}$. For each activated expert $E$, its admissible (replica, server) candidates are formed under the exact-first policy: the exact replicas of $E$ are used whenever at least one is admissible, and a feasible substitute is activated only when no admissible exact candidate remains. The post-compression per-assignment cost of executing $E$ via a replica on $s_n$ from the current residing server $\theta_{i,l}$ is:

$$D_{i,j,l}(E,n) = \frac{\rho_{i,j,l} v_{i,j,l}}{B_{\theta_{i,l},n}} \mathbb{1}[n \neq \theta_{i,l}] + d_{\theta_{i,l},n} \mathbb{1}[n \neq \theta_{i,l}] + (1 - y_{E,n}) \frac{m_E}{\beta_n} \quad (18)$$

which is exactly the post-compression transmission delay (8) plus the GPU-CPU offload delay (2a), measured from $\theta_{i,l}$. Two online guards prune candidates. The hard per-token quality guard:

$$\Lambda_i^{\text{acc}} + \delta_l(\rho_{i,j,l}) \leq \Lambda_i^{max} \quad (19)$$

is never relaxed, and the per-server stability guard

$$W_n(t) + w_E \leq F_n \Delta_t \quad (20)$$

admits a routed expert only if it does not push the load of $s_n$ beyond its capability within the current scheduling window, where $W_n(t)$ is the admitted backlog and $w_E$ is the expert load. The stability guard is the only place where the server capability $F_n$ enters the online decision, consistent with the communication focused scope of Section III.

### *B. Dual-Dimensional Compression and Set-Level Routing*

The distinguishing feature of TopoCompress is that compression and routing are decided jointly per layer under the deployment topology. For each activated token-layer, the router first computes the dual-dimensional score $\mathcal{S}(\tau_{i,j}, l)$ in (6) by a graph lookup, then chooses a compression action, and finally routes the surviving experts as a set. The per-layer subproblem minimizes the post-compression bottleneck cost:

$$J_{i,j,l} = \max_{s_n \in \mathcal{P}_{i,j,l}} \sum_E z_{i,j,l,E,n} D_{i,j,l}(E,n) \quad (21)$$

over the admissible collaboration domain, subject to (19) and (20). Because the $k$ activated experts run in parallel on their participating servers, the layer delay is the bottleneck cost (21) rather than a sum, so two experts on the same server share one fan-out branch while two experts on different servers add a parallel branch. These interactions are invisible to per-expert greedy selection. When the admissible domain is small, the subproblem is solved by exact enumeration of all feasible complete assignments. Otherwise, a beam search of width $W$ is used and refined by 1-exchange local search.

The compression action is chosen by the dual-dimensional score and the budget guard. A token whose score $\mathcal{S}(\tau_{i,j}, l)$ falls below a layer threshold is a candidate for aggressive compression, because it is simultaneously low in semantic value and high in routing cost. The compression ratio $\rho_{i,j,l}$ is then set as small as the budget guard (19) permits. A token whose score is high is preserved at $\rho_{i,j,l} = 1$ to protect both quality and routing efficiency. Similarity-based merging is applied within a group of low-score tokens that share a residing server and a similar representation, fusing them into one representative and setting $\rho_{i,j,l}$ to the retained volume fraction. The continuous $\rho$ produced by the budget guard is rounded to a feasible prune-or-merge action, as defined in Section III.D.

Table II

**Algorithm 2: Online topology-aware compression and collaborative routing**

**Input:** Deployment **X**, residency **Y**, cost graph $\mathcal{G}$ from Algorithm 1; substitute sets; budgets $\{\Lambda_i^{max}\}$; layer thresholds; beam width **W**; slot length $\Delta_t$
**Output:** Compression ratios $\rho$, routing strategy **Z**, token trajectories.
1. for each user $i$ and token $\tau_{i,j}$ do
2. set $\theta_{i,1} \leftarrow a_i$ and $\Lambda_i^{acc} \leftarrow 0$;
3. end for
4. for each layer $l$ = 1 to L do
5. for each active token $\tau_{i,j}$ do
6. obtain $\mathcal{K}_{i,j,l}$ by the gating network
7. look up $\Gamma(E, \theta_{i,l}, \boldsymbol{X})$ from G and compute $\mathcal{C}(\tau_{i,j}, l, \boldsymbol{X})$
8. compute the dual score $\mathcal{S}(\tau_{i,j}, l)$
9. if $\mathcal{S}(\tau_{i,j}, l) < threshold_l$ and budget slack remains then
10. set $\rho_{i,j,l}$ to the smallest value
11. update $\Lambda_i^{acc} += \delta_l(\rho_{i,j,l})$
12. else set $\rho_{i,j,l} \leq 1$
13. end if
14. if the token survives compression then
15. build admissible candidates (exact-first)
16. if admissible set is empty then
17. route each $E$ to its full-precision fallback
18. else solve (21)
19. end if
20. determine $\mathcal{P}_{i,j,l}$ and next residing server $\theta_{i,l+1}$;
21. end if
22. end for
23. accumulate post-compression traffic statistics $\Phi$ for the offline slow loop
24. end for
25. return $\rho$, **Z,** token trajectories.

The Algorithm 2 has three parts: trajectory initialization, per-layer joint compression and routing, and traffic feedback.

**Trajectory initialization (line 1).** Each token $\tau_{i,j}$ is placed on the home server of its user, $\theta_{i,1} \leftarrow a_i$, and its accumulated quality degradation is reset, $\Lambda_i^{acc} \leftarrow 0$. The home server is the only fixed point of the trajectory; from here on the residing server of a token migrates with the experts it is routed to, so that consecutive layers executed on the same server incur no transmission.

**Per-layer joint compression and routing (lines 2-19).** For every active token at layer $l$, the algorithm executes the following steps.

1) **Gating and scoring (lines 4-6).** The gating network produces the Top-k activated set $\mathcal{K}_{i,j,l}$. Using the precomputed graph $\mathcal{G}$, the algorithm looks up the unit-volume routing cost $\Gamma(E, \theta_{i,l}, \boldsymbol{X})$ of each activated expert from the current residing server, aggregates them into $\mathcal{C}(\tau_{i,j}, l, \boldsymbol{X})$ by (5), and combines this with the semantic importance to obtain the dual-dimensional score $\mathcal{S}(\tau_{i,j}, l) = \mathrm{Imp}(\tau_{i,j}, l) \times \mathcal{C}(\tau_{i,j}, l, X)$ of (6). All of this costs $O(k)$ because the routing cost is a table lookup rather than a graph search.

2) **Compression decision under the budget guard (lines 7-11).** If the score is below the layer threshold and the user still has budget slack, the token is a low-value high-cost candidate: the algorithm sets the compression ratio $\rho_{i,j,l}$ to the smallest value that still satisfies the hard quality guard (19), applies the corresponding pruning or similarity-merging action, and adds the incurred degradation $\delta_l(\rho_{i,j,l})$ to $\Lambda_i^{acc}$. Otherwise, the token is preserved at $\rho_{i,j,l} = 1$, contributing zero degradation. This is where the deployment topology actively reshapes the traffic: a token is compressed aggressively precisely when its activated experts are expensive to reach under the current placement.

3) **Candidate construction and pruning (line 13).** For each surviving activated expert, the algorithm builds its admissible candidates under the exact-first policy, considering the feasible substitutes only when no admissible exact replica remains. Candidates that would violate either the per-token quality guard (19) or the per-server stability guard (20) are removed before any cost comparison, so the search operates only over feasible assignments.

4) **Set-level routing (lines 14-16).** If pruning leaves no admissible candidate, the algorithm invokes the emergency fallback and routes each activated expert to its full precision exact replica guaranteed by (c.4). Otherwise, it solves the per-layer subproblem (21) that minimizes the bottleneck cost over the complete collaboration set: when the admissible domain is small it enumerates all feasible complete assignments exactly, and when it is large it runs a width-$W$ beam search refined by 1-exchange local search. Because the objective is the bottleneck (max) cost rather than a sum, the search captures the fan-out and fan-in interactions among co-selected servers that a per-expert greedy choice would miss.

5) **Aggregation and migration (line 17).** Once the $k$ experts are committed, the participating set $\mathcal{P}_{i,j,l}$ is fixed and the next residing server $\theta_{i,l+1}$ is chosen so the token follows the trajectory of its dominant experts. The token is then scattered, executed in parallel, and gathered.

6）**Traffic feedback (lines 20-25).** After all tokens of a layer are processed, the algorithm accumulates the realized post-compression traffic statistics $\Phi$. These statistics are exactly the input that the offline slow loop of Algorithm 1 consumes at the next epoch, which closes the two-timescale alternating loop: the online stage adapts the compression and routing to the current deployment, and the offline stage subsequently re-optimizes the deployment to the traffic that this online stage produced. The procedure repeats across layers and tokens and returns the compression ratios, the routing strategy, and the token trajectories.

It should be note that the TopoCompress alternates between the online fast loop (Algorithm 2, per token-layer) and the offline slow loop (Algorithm 1, per epoch). Let $U(\boldsymbol{X}, \boldsymbol{Y}, \boldsymbol{\rho}, \boldsymbol{Z})$ denote the normalized joint objective (14). In each epoch, the online loop fixes $(\boldsymbol{X}, \boldsymbol{Y})$ and decides $(\boldsymbol{\rho}, \boldsymbol{Z})$ to reduce $U$, and the offline loop then fixes the realized $(\boldsymbol{\rho}, Z)$ statistics and re-optimizes $(\boldsymbol{X}, \boldsymbol{Y})$ to reduce $U$ further. Because each loop only accepts a move that does not increase $U$ on its own variables, the alternation produces a monotonically non-increasing sequence of objective values. Since $U$ is bounded below by zero, the sequence converges to a coordinate wise stationary deployment-compression pair.

*C. Theoretical Properties*

In this section, the properties of the proposed algorithms are investigated.

**Property 1 (Quality feasibility).** Under the full precision

invariant (c.4) and the per-token quality guard (19), Algorithm 2 always produces a quality feasible solution: for every user $i$, $\Lambda_i^{\mathrm{acc}} \leq \Lambda_i^{max}$.

*Proof.* The accumulated degradation is initialized to $\Lambda_i^{\mathrm{acc}} = 0$. It is increased only at two places. First, in the compression step, the ratio $\rho_{i,j,l}$ is chosen as the smallest value satisfying the guard $\Lambda_i^{\mathrm{acc}} + \delta_l(\rho_{i,j,l}) \leq \Lambda_i^{max}$ of (19). Hence, after the update the running degradation still respects the budget. Second, a preserved token sets $\rho_{i,j,l} = 1$, and by the normalization $\delta_l(1) = 0$ of (9) it adds nothing. It remains to check the emergency fallback. When no admissible candidate survives pruning, the fallback routes each activated expert to a full precision exact replica, which exists by the invariant (c.4). Such a replica is full precision, so its quantization loss is 0, and it is the exact expert, so its substitution loss is 0; the total increment to $\Lambda_i^{\mathrm{acc}}$ is therefore 0. Consequently, no step of the algorithm can drive $\Lambda_i^{\mathrm{acc}}$ above $\Lambda_i^{\max}$, and every activated expert receives an assignment, so the routing is complete and quality feasible. ■

**Property 2 (Bounded participating servers).** For every token-layer pair, $|\mathcal{P}_{i,j,l}| \leq k$ regardless of the size of the candidate collaboration domain.

*Proof.* By the routing constraint (c.5), each of the *k* activated experts has exactly one nonzero routing indicator $z_{i,j,l,E,n}$. Therefore, the number of (expert, server) assignments in a layer equals *k*. The participating set $\mathcal{P}_{i,j,l}$ is the set of distinct servers that appear in these *k* assignments, so $|\mathcal{P}_{i,j,l}| \leq k$, with equality when the $k$ experts are placed on *k* distinct servers and $|\mathcal{P}_{i,j,l}| = 1$ when all are co-located. Redundant deployment may enlarge the candidate collaboration domain well beyond *k* servers, but only the servers that are actually selected enter $\mathcal{P}_{i,j,l}$. Hence, the bound is independent of the domain size. This is consistent with the participating-server limit in the problem formulation. ■

**Property 3 (Per-layer routing optimality for small domains).** Let *R* denote the maximum number of admissible (replica, server) candidates of any activated expert after substitution, if the admissible domain size is at most *R* and the compression ratio of the layer is fixed, the exact enumeration in Algorithm 2 returns a global optimum of the per-layer routing subproblem (21) under the current backlog state.

*Proof.* Fix the compression ratio $\rho$ of the layer, so the per-assignment cost $D_{i,j,l}(E,n)$ of (18) is fully determined by the choice of server for each expert under the current backlog. The feasible region of subproblem (21) is the set of complete assignments that pick, for each activated expert, one candidate from its pruned admissible set such that the quality guard (19) and the stability guard (20) both hold. This region is contained in the Cartesian product of the per-expert admissible sets, whose cardinality is at most $R^k$ since each expert has at most *R* candidates. The exact enumeration iterates over every element of this product, discards those violating (19) or (20), evaluates the bottleneck objective in (21) exactly using the true current backlog, and returns the minimizer. Because the search is exhaustive over the entire feasible region and the objective is evaluated without approximation, the returned assignment attains the minimum of (21) over the feasible set, i.e. it is globally optimal for the per-layer subproblem. The optimality is per-layer, not cross-layer: the accumulated budget $\Lambda_i^{\mathrm{acc}}$ and the per-server backlog couple successive layers and tokens, so the layer-wise optima need not compose into a global optimum of P0. ■

**Property 4 (Local optimality of the exchange refinement).** For a large domain, the beam search followed by 1-exchange refinement terminates at a solution that is locally optimal with respect to single-expert reassignment: no feasible reassignment of one activated expert to a different admissible candidate strictly decreases the bottleneck cost (21).

*Proof.* The 1-exchange refinement accepts a single-expert reassignment only if the reassignment is feasible, i.e. it preserves the quality guard (19) and the stability guard (20), and it strictly decreases the bottleneck objective (21). Each accepted move therefore produces a strictly smaller objective value. The feasible region contains at most $R^k$ complete assignments and is thus finite, and the bottleneck objective is bounded below by zero; a strictly decreasing sequence of distinct objective values over a finite set must terminate after finitely many moves. At termination, by the acceptance rule, there exists no feasible single-expert reassignment that strictly decreases (21), which is exactly local optimality over the 1-exchange neighborhood. We do not claim a global approximation ratio: the bottleneck (min-max) structure together with the coupling guards (19) to (20) does not in general admit a constant-factor bound, so the empirical gap to an offline oracle is reported in Section VI instead. ■

**Property 5 (Convergence of the two-timescale alternation).** The alternating optimization between the online fast loop and the offline slow loop produces a monotonically non-increasing sequence of the normalized objective (14) and converges to a coordinate wise stationary deployment compression pair.

*Proof.* Each loop accepts only moves that do not increase $U$ on its own block of variables, so the sequence is non-increasing; since $U \geq 0$, it converges. At the limit, neither re-compressing under the fixed deployment nor re-deploying under the fixed compressed traffic reduces $U$, which is the definition of a coordinate wise stationary point. Global optimality is not claimed because P0 is non-convex.

Let $U(\boldsymbol{X},\boldsymbol{Y},\boldsymbol{\rho},\boldsymbol{Z})$ denote the normalized joint objective (14), and index the epochs by *t*. Within epoch *t*, the online loop fixes the deployment block $(\boldsymbol{X}^t,\boldsymbol{Y}^t)$ and updates the decision block $(\boldsymbol{\rho},\boldsymbol{Z})$. By construction, it only commits a per-token decision when the chosen assignment does not increase the layer contribution to $U$ relative to the feasible alternatives it considers, so $U(\boldsymbol{X}^t,\boldsymbol{Y}^t,\boldsymbol{\rho}^{t+1},\boldsymbol{Z}^{t+1}) \leq U(\boldsymbol{X}^t,\boldsymbol{Y}^t,\boldsymbol{\rho}^t,\boldsymbol{Z}^t)$. The offline loop then fixes the realized $(\boldsymbol{\rho}^{t+1},\boldsymbol{Z}^{t+1})$ statistics and re-optimizes the deployment block. Since every benefit term in Algorithm 1 is an expectation of the same objective $U$ on the post-compression traffic and an insertion or promotion is accepted only when its benefit is positive, $U(\boldsymbol{X}^{t+1},\boldsymbol{Y}^{t+1},\boldsymbol{\rho}^{t+1},\boldsymbol{Z}^{t+1}) \leq U(\boldsymbol{X}^t,\boldsymbol{Y}^t,\boldsymbol{\rho}^{t+1},\boldsymbol{Z}^{t+1})$. Composing the two inequalities, the sequence $\{U^t\}$ is non-increasing across epochs. Since $U \geq 0$ is bounded below, the monotone sequence converges to a limit $U^*$. At the limit neither block can strictly decrease $U$: the online loop cannot reduce $U$ by re-compressing or re-routing under the fixed deployment, and the offline loop cannot reduce $U$ by re-deploying under the fixed compressed traffic. This is a coordinate-wise (block-coordinate) stationary point of $U$. We do not claim global optimality, because P0 is non-convex and block-coordinate descent converges to a stationary point rather than the global minimum. ■

**Property 6 (Online complexity).** The per-token routing cost is $O(LWkR)$ for the beam search variant and $O(LR^k)$ for the exact variant, plus $O(Lk)$ for the dual-score lookups, and is independent of the number of edge servers $N$.

*Proof.* For each layer, computing the dual score and looking up $\Gamma$ over the $k$ activated experts costs $O(k)$ using $\mathcal{G}$. The beam search variant maintains $W$ partial sets and, over $k$ expansion rounds, extends each over at most $R$ candidates, giving $O(WkR)$ per layer and $O(LWkR)$ total. The exact variant has at most $R^k$ complete assignments, each evaluated in $O(k)$, giving $O(LR^k)$. Since Algorithm 1 bounds each replica count by $R$, the overhead is independent of $N$, which makes token-level compression and routing practical for typical Top-$k$ MoE where $k$ is small.

Consider one layer. Computing the dual-dimensional score requires, for each of the $k$ activated experts, one lookup of $\Gamma$ in the precomputed graph $\mathcal{G}$ and a constant amount of arithmetic, costing $O(k)$; over $L$ layers this contributes is $O(Lk)$. For the routing search in the beam search variant, the search maintains at most $W$ partial assignments and proceeds in $k$ expansion rounds. In each round every partial assignment is extended over at most candidate $R$ servers, and each extension evaluates the marginal bottleneck cost in $O(1)$using incrementally maintained branch maxima, giving $O(WkR)$ per layer, and a constant number of 1-exchange sweeps adds the same order.

Over $L$ layers. This is $O(LWkR)$. For the exact variant, the feasible region has at most $R^k$ complete assignments, each evaluated in $O(k)$ for the bottleneck objective, giving $O(R^k)$ per layer and $O(LR^k)$ in total. In all cases the candidate count per expert is bounded by $R$, which Algorithm 1 controls through its replication cap, so the search size depends on $R$ and not on the total number of servers $N$. Hence, the online routing overhead is independent of $N$. Since typical Top-$k$ MoE inference uses a small $k$, both variants remain practical at token-level granularity. ■

## VI. Performance Evaluation

In this section, we evaluate the proposed TopoCompress framework by simulation and analyze its behavior from the perspectives of cross-server communication, latency, token-compression behavior, and inference quality.

### *A. Experimental Setup*

**Experimental setup**. TopoCompress is implemented on PyTorch and evaluated on a trace-driven, geo-distributed edge simulator of 10 edge servers. The servers are heterogeneous in both computation and memory: their computation capacities $F_n$ span 18 to 120 TFLOPS, their GPU memory capacities $G_n^M$ span 12 to 48 GB, and their CPU memory capacities $G_n^C$ span 96 to 512 GB. The servers are interconnected over ordinary Internet links rather than a dedicated fabric, with available bandwidths $B_{m,n}$ ranging from 0.8 to 10 Gbps and propagation-plus-queuing latencies $d_{m,n}$ ranging from 0.3 to 22 ms, so that the bandwidth-limited and time-varying nature of edge interconnections is reproduced. Each user is attached to one access edge server following a non-uniform spatial distribution, and requests arrive at a default rate of 40 requests per second unless stated otherwise.

**Models, tasks, and calibration**. We evaluate TopoCompress on three MoE models of increasing scale, Switch-Base-8E, Qwen-MoE-A2.7B, and Mixtral-8x7B. Inference quality is measured on WikiText-103 (perplexity), SQuAD (F1), and GSM8K (accuracy). All offline quantities required by Algorithm 1 (the expert deployment, the GPU-CPU residency, the routing-cost graph, the per-layer compression thresholds, and the post-compression traffic statistics) are computed once on a held-out calibration set with the backbone strictly frozen. Unless otherwise stated, the per-token quality-degradation budget is set to 2%, the beam width of the set-level router is $W = 8$, the memory budget ratio of the redundant deployment is 2.0, and the two-timescale alternation is run for 5 epochs.

**Baselines.** We compare TopoCompress with four representative methods that cover the relevant design space: 1) DanceMoE [19], a distributed edge MoE scheme that optimizes the expert placement without compressing the token stream; 2) MoE-Infinity [18], an activation-aware expert offloading and caching scheme; 3) EdgeShard [23], a collaborative edge LLM inference scheme based on dense model partitioning; and 4) SnapKV [27], a topology-agnostic token-compression scheme that prunes tokens purely by their semantic importance. We further include the classic token-compression method H2O [29] as a second topology-agnostic compression reference. For the ablation study, we include four internal variants, i.e., TopoCompress without the topology-aware score (compression by semantic importance only), without the GPU-CPU residency optimization, without the traffic-aware redundant replication, and without the offline slow loop (one-directional compression only). The metrics are the cross-server communication volume, the average and tail (P99) inference latency, the remote execution ratio, the token-compression ratio, the throughput, and the inference quality.

### *B. Overall Performance Comparison*

We first compare the overall performance on Mixtral-8x7B, as shown in Fig. 2, which reports the cross-server traffic per one thousand tokens, the average latency, the remote execution ratio, and the throughput.

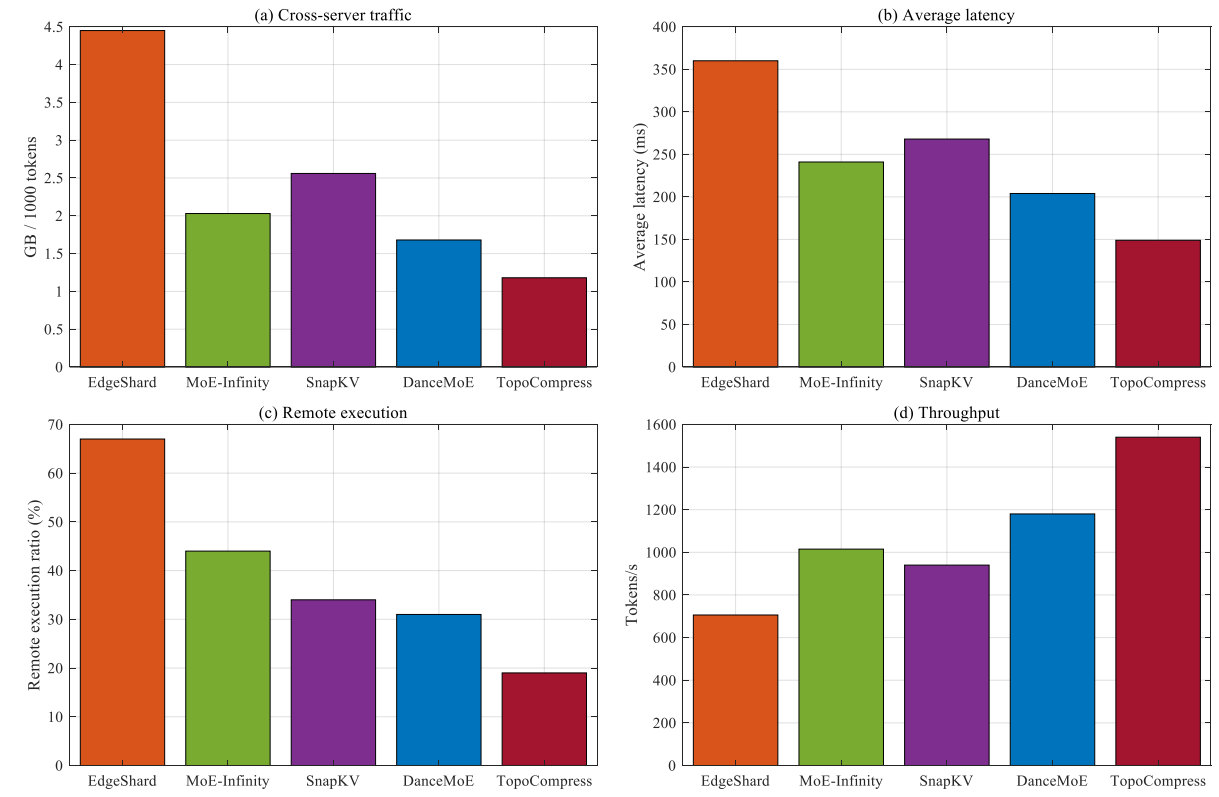

Fig. 2 Overall performance

As shown in Fig. 2(a), TopoCompress attains the lowest cross-server traffic of 1.18 GB per one thousand tokens, which is 73.5% lower than EdgeShard, 41.9% lower than MoE-Infinity, and 29.8% lower than DanceMoE. The additional gain over DanceMoE is the direct effect of the deployment-topology-aware token compression: a token that is both semantically unimportant and expensive to route is compressed on the spot, so its representation is never transmitted across servers, whereas DanceMoE optimizes only the placement and still transmits the full token stream. Fig. 2(b) shows that the average latency of TopoCompress is only 149 ms, 58.6% below EdgeShard and 27.0% below DanceMoE, because the saved cross-server transmission directly shortens the per-layer fan-out and fan-in delay on the bandwidth-limited edge links. Fig. 2(c) shows that its remote execution ratio drops to 19%, far below the 67% of EdgeShard and the 31% of DanceMoE, since compressing the high-cost tokens removes exactly the assignments that would otherwise cross servers. Consistently, Fig. 2(d) shows that TopoCompress delivers the highest throughput of 1540 tokens per second, 2.18x that of EdgeShard.

It is worth noting that SnapKV reduces traffic relative to EdgeShard but remains clearly worse than TopoCompress, because it compresses tokens without being aware of which executions are cross-server, so it often drops cheap local tokens while keeping the expensive remote ones.

### C. Tail Latency and Compression Behavior

Fig. 3 reports the latency distribution and the token execution-type breakdown. Fig. 3(a) plots the empirical CDF of the per-request latency, and Fig. 3(b) decomposes all token-layer executions into local exact, remote exact, GPU-CPU offloaded, compressed (pruned), compressed (merged), and emergency fallback.

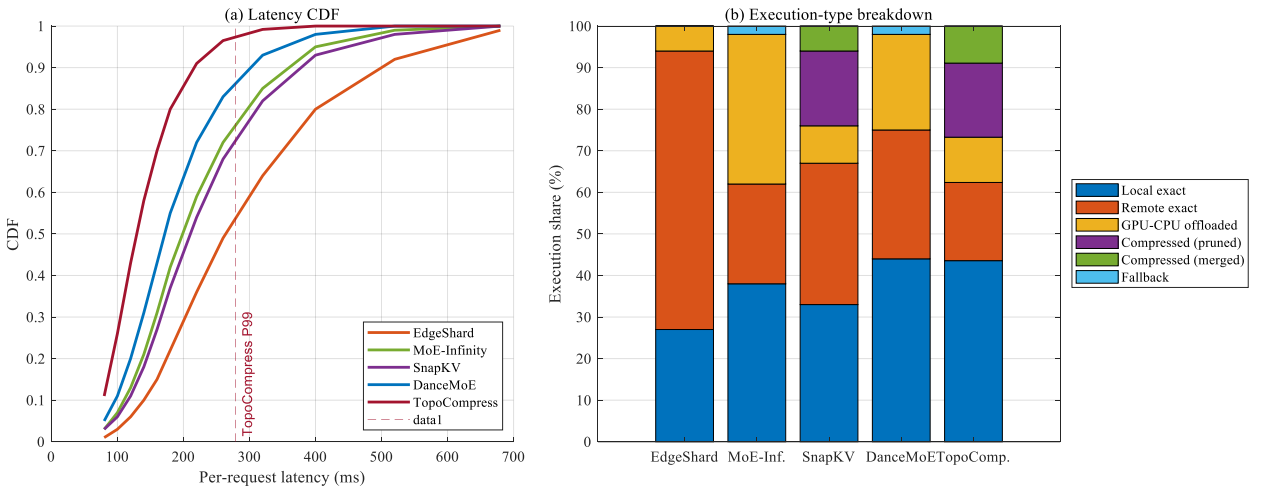

Fig. 3 Tail latency and compression behavior

From Fig. 3(a), TopoCompress not only shifts the whole distribution to the left but also compresses the tail: its P99 latency is 279 ms, 35.7% lower than that of DanceMoE and 61.8% lower than that of EdgeShard. The compressed tail comes from the set-level routing together with the topology-aware compression, which avoids the worst-case multi-hop trajectories by both removing the high-cost tokens and routing the surviving experts as a set rather than greedily. From Fig. 3(b), TopoCompress compresses 27% of the token-layer executions entirely, 18% pruned and 9% merged, and among the remaining executions only 19% are remote and only 11% require GPU-CPU offloading. In contrast, EdgeShard transmits every token exactly and incurs 67% remote executions, while SnapKV compresses tokens but, being topology-agnostic, still leaves 34% remote executions. This confirms that TopoCompress attacks the cross-server traffic at its source, which a topology-agnostic compressor alone cannot remove.

### D. Inference Quality and the Quality Budget

Fig. 4 evaluates the inference quality. Fig. 4(a) reports the quality of representative methods on the three datasets, and Fig. 4(b) shows how the cross-server traffic and the actual quality degradation of TopoCompress vary with the per-token quality budget.

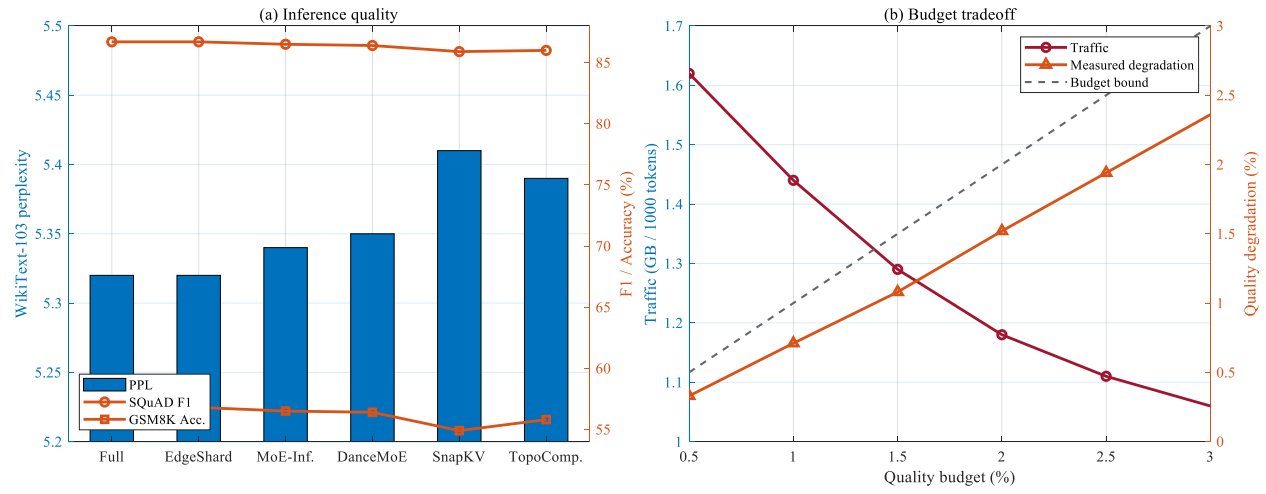

Fig. 4 Inference quality and quality budget

From Fig. 4(a), the quality of TopoCompress stays close to the uncompressed methods: compared with the full centralized model, the perplexity on WikiText-103 rises by only 0.34, the SQuAD F1 drops by only 0.7, and the GSM8K accuracy drops by only 1.0. This small and controlled loss is a consequence of the design: the compression of a token is admitted only when its dual-dimensional score is low and the per-token quality guard still has slack, so high-value tokens are always preserved at full precision. By comparison, SnapKV compresses tokens by semantic importance alone and, without a per-token budget tied to the routing cost, suffers a larger GSM8K drop of 1.9. From Fig. 4(b), as the quality budget is relaxed from 0.5% to 3%, TopoCompress trades quality for communication in a smooth and monotone manner, and the measured degradation always stays strictly below the budget bound, which is exactly the hard guarantee established in Property 1. The knee of the curve is around a 2% budget, where the traffic is already close to its minimum while the degradation is still well controlled, which justifies the default setting.

### E. Token-Compression Behavior

Fig. 5 examines how TopoCompress allocates compression. Fig. 5(a) reports the per-layer compression ratio and the average per-layer routing cost on Mixtral-8x7B, and Fig. 5(b) reports the joint distribution of the compressed tokens over the two score dimensions, i.e., the semantic importance and the topology-induced routing cost.

From Fig. 5(a), the compression ratio is low in the shallow layers, where the token representations are not yet stabilized and are quality-sensitive, and rises in the middle-to-deep layers, where the per-layer routing cost is also higher because the experts are more widely distributed; the two curves are positively correlated, which shows that TopoCompress compresses more exactly where the cross-server cost is larger. From Fig. 5(b), the compressed tokens concentrate in the low-importance, high-routing-cost quadrant, while the high-importance tokens and the low-cost local tokens are largely preserved. This is precisely the behavior intended by the dual-dimensional score in (6): a token is compressed only when it is simultaneously cheap to drop in quality and expensive to keep in communication, which explains why the quality loss remains bounded even though more than a quarter of the token-layer executions are removed.

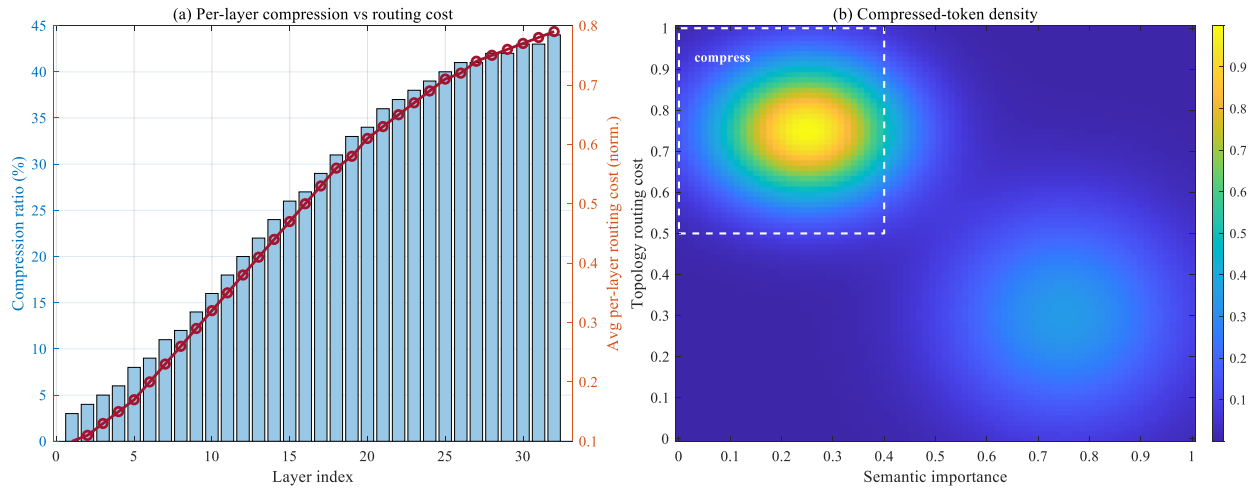

Fig. 5 Token-compression behavior

### F. Effect of the Two-Timescale Alternation

Fig. 6 evaluates the effect of the offline-online alternation. Fig. 6(a) shows the normalized joint objective over the alternation epochs, and Fig. 6(b) shows the cross-server traffic and the CPU-offload ratio over the epochs.

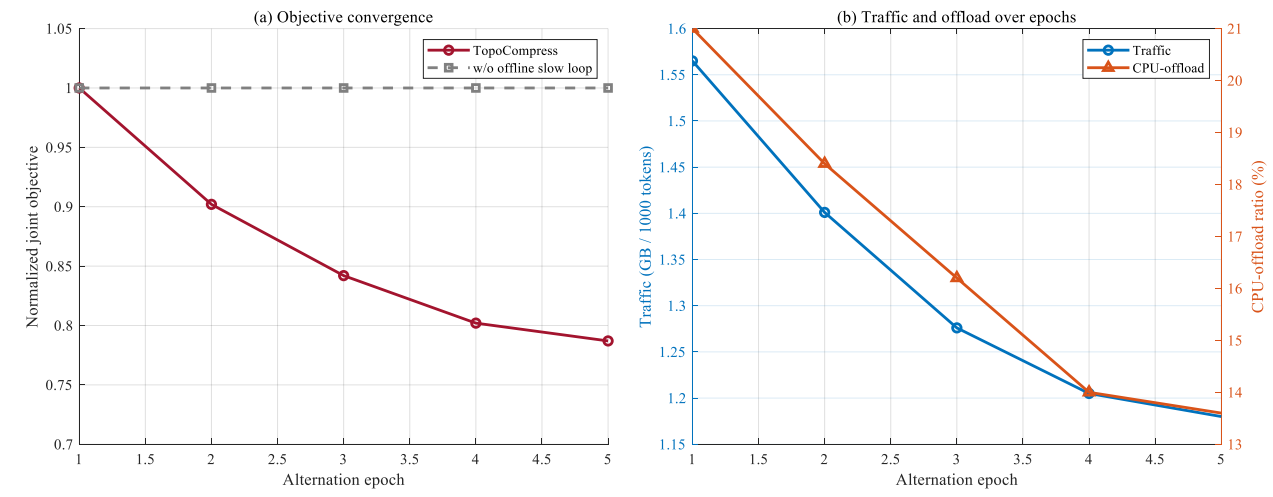

Fig. 6 Effect of the two-timescale alternation

As shown in Fig. 6(a), the normalized joint objective decreases monotonically over the epochs and stabilizes within 4 to 5 epochs, by 21.3% relative to the first epoch, which empirically confirms the convergence established in Property 5: each epoch the online loop adapts the compression to the current deployment, and the offline loop re-optimizes the deployment to the resulting post-compression traffic, so the objective cannot increase. From Fig. 6(b), the cross-server traffic and the CPU-offload ratio both drop quickly in the first three epochs, by 24.6% and 19.0% respectively, and then saturate, because the offline loop reclaims the redundant replicas whose traffic has been compressed away and promotes into GPU memory the replicas that still serve the surviving high-cost traffic. A one-directional scheme that fixes the deployment and only adapts the compression, i.e., the "without offline slow loop" variant, stops at the first-epoch level and is clearly worse, which shows that the reverse coupling from compression to deployment is a real source of gain rather than a redundant module.

### G. Comparison Across MoE Models

Fig. 7 compares TopoCompress with the representative baselines on the three MoE models. Fig. 7(a) reports the cross-server traffic, and Fig. 7(b) reports the quality retention relative to the full centralized model.

As shown in Fig. 7(a), TopoCompress consistently achieves the lowest cross-server traffic, and its advantage widens as the model scales: the traffic reduction relative to DanceMoE grows from 26.1% on Switch-Base-8E to 29.8% on Mixtral-8x7B, while the absolute saving grows from 0.34 GB to 0.50 GB per one thousand tokens. Larger models have more layers and more experts, which both increases the number of cross-server-prone tokens that the topology-aware compression can remove and enlarges the heterogeneity of the routing cost that the dual-dimensional score can exploit. As shown in Fig. 7(b), the quality retention of TopoCompress stays above 98.8% on all three models, confirming that the communication gains do not come at the cost of inference quality.

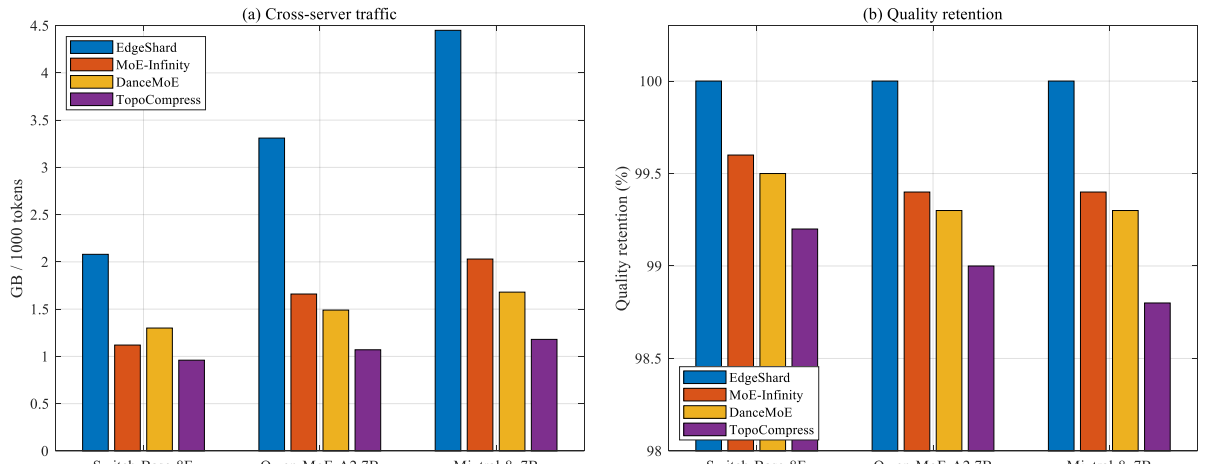

Fig. 7 Comparison across MoE models

### H. Ablation Study

Fig. 8 isolates the contribution of each component on Mixtral-8x7B by removing the topology-aware score, the GPU-CPU residency optimization, the traffic-aware redundant replication, and the offline slow loop in turn. Fig. 8(a) and Fig. 8(b) report the cross-server traffic and the average latency, and Fig. 8(c) reports the actual quality degradation.

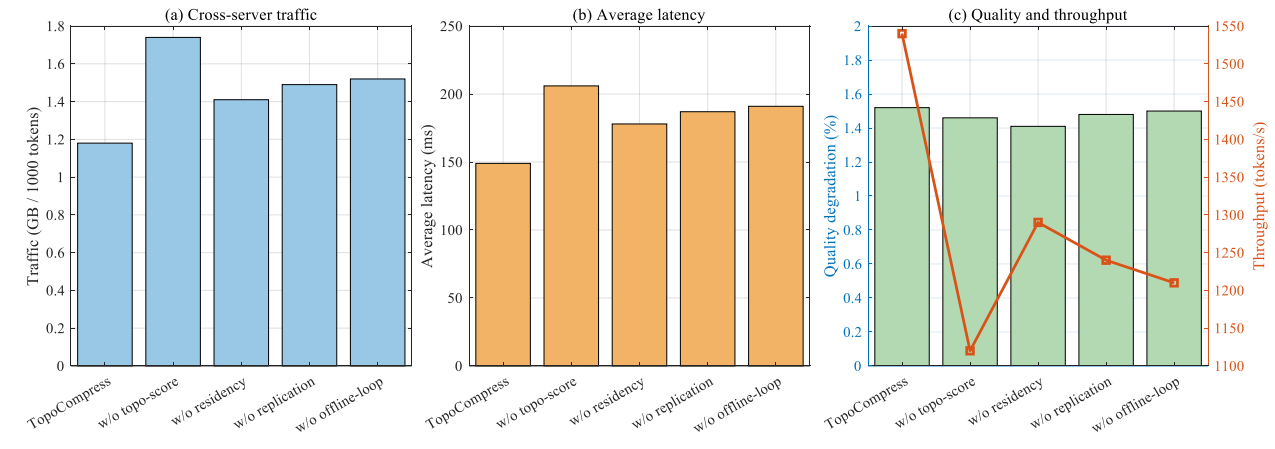

Fig. 8 Ablation study

From Fig. 8(a) and Fig. 8(b), removing any component increases both traffic and latency. Disabling the topology-aware score causes the largest traffic increase, from 1.18 to 1.74 GB per one thousand tokens, because the compressor then drops tokens by semantic importance alone and no longer targets the high-cost cross-server tokens; this is the component that most directly realizes the central idea of the paper. Disabling the offline slow loop increases the traffic to 1.52 GB, because the deployment can no longer be re-optimized to the compressed traffic, which confirms that the bidirectional coupling is necessary. Disabling the GPU-CPU residency optimization and the traffic-aware replication also degrade performance, though to a smaller extent. From Fig. 8(c), the quality degradation of all variants stays small and even decreases slightly when the topology-aware score is removed, which shows that the topology-aware components mainly serve to cut communication rather than to preserve quality, since the quality is already protected by the per-token budget regardless of which component is enabled.

### I. Performance Under Different Request Loads

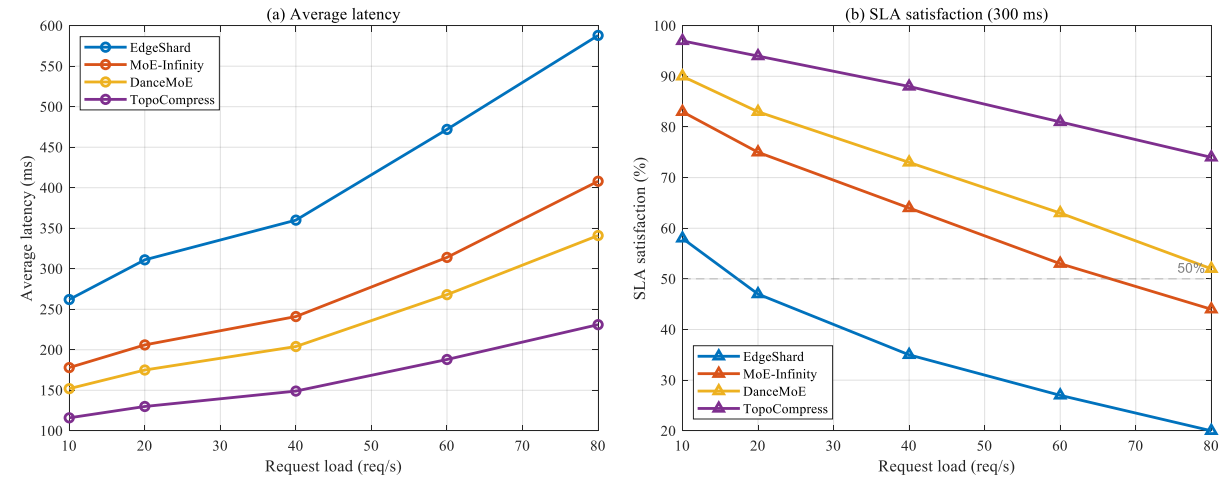

Fig.9 Robustness under different request loads

Finally, Fig. 9 evaluates robustness to the request load, varied from 10 to 80 requests per second.

As shown in Fig. 9(a), the average latency of all methods grows with the load because of increased GPU queueing and link contention, but TopoCompress grows the slowest, since fewer and shorter cross-server transmissions relieve both the busy links and the GPU queues. As shown in Fig. 9(b), where the SLA satisfaction ratio is defined as the fraction of requests completed below a 300 ms target, TopoCompress still satisfies 74% of requests at 80 requests per second, whereas the baselines fall below 52% under the same load. This confirms that, by removing the high-cost tokens and the transmissions they trigger, TopoCompress remains markedly more robust under heavy load.

## VII. Conclusion

In this paper, we identify a gap in distributed edge MoE inference research: existing work focuses on expert placement, caching, replication, offloading, and communication scheduling for raw token traffic, but neglects deployment- and topology-aware token compression. To reduce cross-server transmission while balancing inference quality and resource consumption, we propose TopoCompress, a compression framework that jointly considers token semantic importance, topology-dependent routing costs, expert deployment and replication, GPU–CPU residency, and collaborative routing. We formulate a weight-based joint optimization to decide which tokens to compress and how to route surviving expert activations. To handle the coupling between per-token compression and per-epoch deployment, we design a two-timescale alternating optimization: a fast online loop compresses semantically unimportant but communication-expensive tokens under quality budgets and routes surviving activations collaboratively, while a slow offline loop re-optimizes expert placement, replication, and residency based on post-compression traffic. We also analyze feasibility, optimality, convergence, and complexity. Simulations show TopoCompress significantly reduces cross-server traffic and deployment cost with small, controllable quality loss.

## Acknowledgment

This work was supported in part by the grant from NSFC Grant no. 62571156, 62101159, NSF of Shandong Grant no. ZR2021MF055, the Research Grants Council of Hong Kong under the Areas of Excellence scheme grant AoE/E-601/22-R.

Additionally, the authors used an AI-based language assistance tool to improve the clarity and readability of parts of the manuscript, particularly the Abstract and Introduction, and all technical content, analysis, and conclusions were developed and carefully verified by the authors.